\documentclass[twocolumn,superscriptaddress,aps,preprintnumbers,amsmath,amssymb,sort&compress,nofootinbib
]{revtex4-1}
\usepackage{amsfonts,amsmath,amssymb,ascmac,bm,tensor}
\usepackage{comment}
\usepackage{ifpdf}
\usepackage{slashed}
\usepackage[dvipdfmx]{color}
\usepackage[mathscr]{eucal}
\usepackage[utopia]{mathdesign}
\usepackage[utf8]{inputenc}

\usepackage{cancel}

\ifpdf
  \usepackage{graphicx}     %   usepackage without driver option
  \usepackage[bookmarksopen,colorlinks=true,linkcolor=bblue,citecolor=bblue,urlcolor=ppink]{hyperref}
\else     % For (p)LaTeX + dvipdfmx
  \usepackage[dvipdfmx]{graphicx}     %   usepackage with driver option
  \usepackage[dvipdfmx,bookmarksopen,colorlinks=true,linkcolor=bblue,citecolor=ppink,urlcolor=darkred]{hyperref}
\fi

\definecolor{red}{rgb}{1,0,0}
\definecolor{darkred}{rgb}{0.6,0,0}
\definecolor{darkgreen}{rgb}{0.992447,0.623778,0.034597}
\definecolor{ppink}{rgb}{1,0.4,0.4}
\definecolor{bblue}{rgb}{0.284602,0.317763,0.963947}
\newcommand{\1}{\mbox{1}\hspace{-0.25em}\mbox{l}}
\newcommand{\vev}[1]{ \left< {#1} \right> }

\newcommand{\prn}[1]{\left( {#1} \right)}

\newcommand{\dd}{\mathrm{d}}

\newcommand{\ee}{\mathrm{e}}

\makeatletter
\newcommand\footnoteref[1]{\protected@xdef\@thefnmark{\ref{#1}}\@footnotemark}
\makeatother

\begin{document}

%%%%%%%%%%%%%%%%%%%%%%%%%%%
%%%%%%%%%%% Title %%%%%%%%%%%
%%%%%%%%%%%%%%%%%%%%%%%%%%%

%%paper
\title{
Primordial black holes and uncertainties in the choice of the window function
}
\author{Kenta Ando}
\affiliation{ICRR, University of Tokyo, Kashiwa, 277-8582, Japan}
\affiliation{Kavli IPMU (WPI), UTIAS, University of Tokyo, Kashiwa, 277-8583, Japan}
\author{Keisuke Inomata}
\affiliation{ICRR, University of Tokyo, Kashiwa, 277-8582, Japan}
\affiliation{Kavli IPMU (WPI), UTIAS, University of Tokyo, Kashiwa, 277-8583, Japan}
\author{Masahiro Kawasaki}
\affiliation{ICRR, University of Tokyo, Kashiwa, 277-8582, Japan}
\affiliation{Kavli IPMU (WPI), UTIAS, University of Tokyo, Kashiwa, 277-8583, Japan}

\begin{abstract}

\noindent
Primordial black holes (PBHs) can be produced by the perturbations that exit the horizon during inflationary phase.
While inflation models predict the power spectrum of the perturbations in Fourier space,
the PBH abundance depends on the probability distribution function (PDF) of density perturbations in real space.
In order to estimate the PBH abundance in a given inflation model, we must relate the power spectrum in Fourier space to the PDF in real space by coarse-graining the perturbations with a window function.
However, there are uncertainties on what window function should be used, which could change the relation between the PBH abundance and the power spectrum.
This is particularly important in considering PBHs with mass $30 M_\odot$ that account for the LIGO events because the required power spectrum is severely constrained by the observations.
In this paper, we investigate how large influence the uncertainties on the choice of a window function have over the power spectrum required for LIGO PBHs.
As a result, it is found that the uncertainties significantly affect the prediction for the stochastic gravitational waves (GWs) induced by the second order effect of the perturbations. In particular, the pulsar timing array constraints on the produced GWs could disappear for the real-space top-hat window function.

\end{abstract}

\date{\today}
\maketitle
\preprint{IPMU 18-0033}
%\preprint{DESY 18-???}

%%%%%%%%%%%%%%%%%%%%%%%%%%%%%%%%
%%%%%%%%%%% Introduction %%%%%%%%%%%
%%%%%%%%%%%%%%%%%%%%%%%%%%%%%%%%%
\section{Introduction}\label{sec: intro}

LIGO-Virgo Collaboration has detected several events of gravitational waves (GWs) that are produced through mergers of black holes (BHs) or neutron star(s)~\cite{Abbott:2016blz,Abbott:2016nmj,Abbott:2017vtc,Abbott:2017oio,TheLIGOScientific:2017qsa,Abbott:2017gyy}.
Some events are caused by the mergers of BHs whose masses are about $30M_\odot$ (GW150914~\cite{Abbott:2016blz}, GW170104~\cite{Abbott:2017vtc}, GW170814~\cite{Abbott:2017oio}),
 which might be too heavy for the stellar BHs produced in the usual metallicity environment ($Z\sim Z_\odot$)~\cite{TheLIGOScientific:2016htt,Belczynski:2009xy,Spera:2015vkd}.\footnote{
BHs produced in a low-metallicity environment are one of the candidates for $30M_\odot$ BHs detected by LIGO~\cite{Kinugawa:2014zha}.
 }
 On the other hand, primordial black holes (PBHs)~\cite{Hawking:1971ei,Carr:1974nx,Carr:1975qj} can be $30M_\odot$ because the mass of PBHs is determined by the scale of perturbations producing PBHs.
Therefore, PBHs are good candidates for the $30M_\odot$ BHs detected by LIGO~\cite{Bird:2016dcv,Clesse:2016vqa,Sasaki:2016jop,Eroshenko:2016hmn,Carr:2016drx}.

PBHs can be produced by the large perturbations that exit horizon during inflation~\cite{GarciaBellido:1996qt,Kawasaki:1997ju,Yokoyama:1998pt}.
If we determine an inflation model, we can predict the power spectrum of the perturbations in Fourier space.
On the other hand, the PBH abundance is determined by the probability distribution function (PDF) of perturbations in real space.
In order to relate the power spectrum in Fourier space to the PDF in real space,
we must apply the coarse-graining procedure with window functions.
Despite the existence of some window functions, there is not a broad consensus on what window function should be used.
Moreover, a different choice of a window function leads to a different relation between the PBH abundance and the required power spectrum,
though, in realistic situations, the PBH abundance should have a one-to-one correspondence with the power spectrum.
This causes an uncertainty in the estimation of the PBH abundance for a given power spectrum.
In particular, in the context of the PBHs for LIGO events, the relation is essential for predicting the observable quantities such as the $\mu$-distortion~\cite{Fixsen:1996nj,Chluba:2012we,Kohri:2014lza}  and the stochastic GWs~\cite{Saito:2008jc,Saito:2009jt,Bugaev:2009zh,Bugaev:2010bb,Inomata:2016rbd,Orlofsky:2016vbd} produced by the perturbations required for LIGO PBH formation.
The perturbations producing $30M_\odot$ PBHs, whose scales are $k\sim10^{6}$\,Mpc$^{-1}$, cause the $\mu$-distortion in the cosmic microwave background (CMB) spectrum and the stochastic GWs from their second order effect.
The produced $\mu$-distortion and stochastic GWs are constrained by COBE/FIRAS~\cite{Fixsen:1996nj} and the pulsar timing array (PTA) experiments~\cite{Lentati:2015qwp,Shannon:2015ect,Arzoumanian:2015liz} respectively.
In the previous paper~\cite{Inomata:2016rbd}, the Gaussian window function was used and it was shown that the peak of the power spectrum at $k\sim10^{6}$\,Mpc$^{-1}$ must be rapidly damped on both the larger and the smaller scales to avoid $\mu$-distortion and the PTA constraints.
This means that, in the case with the Gaussian window function, the PBH mass spectra must have a sharp peak around $30M_\odot$ and cannot extend to the lighter mass range such as $\mathcal O(1)M_\odot$ or the heavier mass range such as $\mathcal O(1000) M_\odot$.
Since LIGO has a potential to detect GWs produced by the mergers of such light or heavy BHs,
 it is important to make clear how much the power spectra required for LIGO PBHs change depending on the choice of a window function.

In this paper, we take three commonly used window functions, the real-space top-hat window function, Gaussian window function, and Fourier (k)-space top-hat window function as concrete examples.
We investigate the uncertainties on the observable quantities originating from those on the choice of a window function by calculating the necessary perturbations for LIGO PBHs and predicted observable quantities with each window function.

%%%%%%%%%%%%%%%%%%%%%%%%%%%%%%%%%%%%
\section{Formulae for PBH formation}
\label{sec:pbh_formation}
%%%%%%%%%%%%%%%%%%%%%%%%%%%%%%%%%%%%

In this section, we summarize the basic formulae for PBH formation.

In this paper, we focus on PBHs produced during the radiation-dominated era.\footnote{
PBHs produced during the matter-dominated era are discussed in Refs.~\cite{Khlopov:1980mg,Khlopov:1982sov,Harada:2016mhb}.
}
When sufficiently large perturbations reenter the horizon, the gravity of the over-dense regions can overcome the radiation pressure and collapse to PBHs.
The threshold for the PBH formation has been originally estimated with a simple analysis as $\delta_c = 1/3$~\cite{Carr:1975qj} and then 
numerically calculated by several authors as $0.4 \lesssim \delta_c \lesssim 0.6$~\cite{Musco:2004ak, Polnarev:2006aa, Musco:2012au, Harada:2015yda},
where $\delta_c$ is the threshold of the density perturbations in the comoving gauge at the horizon reentry.
We stress again that although the authors of the numerical simulations assume density profiles in the real space and take the real-space top-hat window function when they relate the density profile to the threshold~\cite{Musco:2004ak, Polnarev:2006aa, Musco:2012au, Harada:2015yda},
it is not clear what window function we should take when we relate the power-spectra in the Fourier space to the PDF.
We conservatively take $\delta_c=0.4$ as a fiducial value in the following.\footnote{%%
If $\delta_c=0.6$ is taken instead, the amplitude required for the LIGO PBHs becomes larger and constraints from $\mu$-distortion and PTA observations become more severe.
} 

The mass of a PBH is nearly equal to the horizon mass at the horizon reentry of the perturbation.
The mass is related to the scale of the perturbation as
\begin{align}
	M
	&= \left. \gamma \rho \frac{4 \pi H^{-3}}{3} \right|_{k = aH}
	\simeq\frac{\gamma M_\text{eq}}{\sqrt{2}}
	\prn{ \frac{g_{\ast,\text{eq}}}{g_\ast} }^\frac{1}{6}
	\prn{ \frac{k_\text{eq}}{k} }^2  \nonumber \\[.5em]
	& \simeq \gamma M_\odot
	\left(\frac{g_*}{10.75}\right)^{-\frac{1}{6}}\left(\frac{k}{4.2\times10^6\,\mathrm{Mpc}^{-1}}\right)^{-2}
%	\left(\frac{k}{1.9\times10^6\,\mathrm{Mpc}^{-1}}\right)^{-2}
	\label{eq:pbhmass in k} \\[.5em]
	&\simeq \gamma M_\odot
    \left(
    \frac{g_\ast}{10.75}
    \right)^{- \frac{1}{6}}
    \left(
        \frac{f}{6.5 \times 10^{-9} \,\textrm{Hz}}
%    \frac{f}{2.9 \times 10^{-9} \,\textrm{Hz}}
    \right)^{-2} \label{eq:pbhmass_freq},
\end{align}
where the corresponding frequency,  $f\equiv k/2\pi$, has been derived for later convenience.
The subscript "eq" means the value at the matter-radiation equality time.
For example, $M_\text{eq}$ is the horizon mass at the equality time.
$g_*$ ($g_{*,\text{eq}}$) is the effective number of relativistic degrees of freedom contributing to the radiation energy density at the PBH formation (at the equality time).
$\gamma$ indicates the ratio of the PBH mass to the horizon mass at the horizon reentry.
Although the value of $\gamma$ is estimated as $\gamma \simeq 0.2$ with the simple analysis~\cite{Carr:1975qj},
$\gamma$ depends on the detail of the gravitational collapse and still has uncertainties.
Therefore, in addition to the case with $\gamma=0.2$, we also consider the cases with $\gamma=1$ for a conservative discussion in this paper.

If the perturbations follow the Gaussian PDF\footnote{
The case where there are non-Gaussianities is discussed in Refs.~\cite{Byrnes:2012yx,Young:2013oia,Nakama:2016kfq,Nakama:2016gzw,Nakama:2017xvq}.
}, the PBH production rate $\beta(M)$ is given by~\cite{Carr:1975qj}\footnote{%%
More precisely $\beta$ is the production rate per logarithmic Hubble time interval, i.e. $\dd \ln H^{-1}$ ($= \dd \ln M$).}:
\begin{align}
	\beta (M) =
	\int_{\delta_c}
	\frac{\dd \delta}{\sqrt{2 \pi \sigma^2 (M)}} \, \ee^{- \frac{\delta^2}{2 \sigma^2 (M)}}
	\simeq
	\frac{1}{\sqrt{2 \pi}} \frac{1}{\delta_c / \sigma (M)} \, \ee^{- \frac{\delta_c^2}{2 \sigma^2 (M)}}.
	\label{eq:beta}
\end{align}
$\sigma^2(M)$ is the mean square of coarse-grained density perturbations with the smoothing scale $k^{-1}$ at the horizon reentry, which is given by~\cite{Liddle:452061}
\begin{align}
	\sigma^2 (M (k))
	= \vev{\delta^2(R=k^{-1},\bm x, \eta = k^{-1}) },
\end{align}
where $R$ is the smoothing scale and $\eta$ is the conformal time.
$\delta(R,x,\eta)$ is the coarse-grained density perturbations at $\eta$, which is defined with a window function $W(R, x)$ as
\begin{align}
\delta(R,\bm x, \eta) \equiv \int W(R, |\bm x - \bm x'|) \delta(\bm x' , \eta) \dd^3 x'.
\end{align}
The Fourier component of $\delta (R, \bm x, \eta)$ is given by
\begin{align}
	\delta (R, \bm k, \eta) &\equiv \int \dd^3 x \delta(R,\bm x, \eta) \ee^{-i\bm k \cdot \bm x} \nonumber \\
	& = \tilde W(R, k) \delta(\bm k, \eta),
\end{align}
where $\delta (\bm k, \eta)$ and $\tilde W(R,k)$ are the Fourier components of $\delta (\bm x, \eta)$ and $W(R,x)$.
Then, we finally get~\cite{Blais:2002gw,Josan:2009qn}
\begin{align}
 	\sigma^2 (M (k))
	&= \int \dd \ln q \tilde W^2 (k^{-1}, q) \mathcal P_\delta(q, \eta=k^{-1}) \nonumber \\
	 &= \int \dd \ln q \tilde W^2 (k^{-1}, q) \frac{16}{81} \prn{q k^{-1}}^4 \nonumber \\
	 &\qquad \qquad \qquad \qquad
	 T^2 (q, \eta = k^{-1}) \mathcal P_\mathcal R (q),
	\label{eq:sigma}
\end{align}
where $T(k, \eta)$ is the transfer function defined as
\begin{align}
	T(k, \eta) = 3 \frac{\sin(k\eta/\sqrt{3}) - (k\eta/\sqrt{3}) \cos(k\eta/\sqrt{3}) }{(k\eta/\sqrt{3})^3}.
	\label{eq:transfer_formula}
\end{align}
The transfer function describes the evolution of the subhorizon modes.
The expression for the transfer function in Eq.~(\ref{eq:transfer_formula}) is valid only during the radiation-dominated era, which is the case we consider here.\footnote{
Note that if we take the Gaussian or the k-space top-hat window function, the transfer function is not important.
This is because, in the case with the two window functions, the density perturbations coarse-grained with the horizon scale are insensitive to the subhorizon modes.
}
$\mathcal P_\delta(k,\eta)$ and $\mathcal P_\mathcal R(k)$ are the power spectra of the density perturbations and curvature perturbations,
where $\mathcal P_\mathcal R(k)$ is the power spectrum in the superhorizon limit.
Although it is not clear what window function should be used,
the value of $\beta(M)$ significantly depends on the shape of the window function.
This is the main point of this paper.

The current fraction of dark matter (DM) in PBHs is given by
\begin{align}
	&f(M)
	\simeq  \left. \frac{\rho_\text{PBH}(M)}{\rho_m} \right|_\text{eq} \frac{\Omega_{m}}{\Omega_\text{DM}}
	= \prn{\frac{T_M}{T_\text{eq}} \frac{\Omega_{m}}{\Omega_\text{DM}}} \gamma \beta (M) \nonumber \\
	& \simeq
	\gamma^{\frac{3}{2}}
	\!\left(\! \frac{\beta (M)}{1.6 \times 10^{-9}} \!\right)
	\!\left(\! \frac{10.75}{g_{\ast} (T_M)} \!\right)^\frac{1}{4}\!
	\!\left(\! \frac{0.12}{\Omega_\text{DM} h^2} \!\right)
	\!\left(\! \frac{M}{M_\odot} \!\right)^{-\frac{1}{2}}, \hspace{-7pt} \label{eq:frac1}
\end{align}
where $f(M) \equiv \frac{1}{\Omega_\text{DM}} \frac{\dd\, \Omega_\text{PBH}}{\dd\, \text{ln}\,M}$ and $\rho_\text{PBH}(M)\equiv \frac{\dd\, \rho_\text{PBH}}{ \dd\, \text{ln} \, M}$
 are the differential mass function of the PBH DM fraction and the PBH energy density, respectively.
The subscripts ``$m$'' and ``DM'' mean the matter (baryon + DM) and DM (DM only), with $\Omega_\text{DM} h^2 \simeq 0.12$~\cite{Ade:2015xua}.
$T_M$ represents the temperature at which the PBHs with mass $M$ are produced.
Since $f(M)$ is the differential mass function, the total fraction of DM in PBHs is given by
\begin{align}
	\frac{\Omega_{\text{PBH,tot}}}{\Omega_\text{DM}} = \int \dd \ln M\, f(M).
\end{align}
%%

%%%%%%%%%%%%%%%%%%%%%%%%%%%%%%%%%%%%
\section{Properties of window functions}
\label{sec:window}
%%%%%%%%%%%%%%%%%%%%%%%%%%%%%%%%%%%%

In this section, we discuss the properties of the window functions.
We take the real-space top-hat window function, Gaussian window function, and k-space top-hat window function as concrete examples.
The window functions are given as follows,

\vspace{0.2em}
\noindent \emph{real-space top-hat window function:}
\begin{align}
	W(R,x) &= \left( \frac{4\pi}{3} R^3 \right)^{-1} \Theta(R-x), \\[.5em]
	\tilde W(R,k) &= 3\left( \frac{\sin(kR) - kR \cos(kR)}{(kR)^3} \right),
\end{align}

\noindent \emph{Gaussian window function:}
\begin{align}
	W(R,x) &= \left( (2\pi)^{3/2}R^3  \right)^{-1} \exp\left(-\frac{x^2}{2R^2}\right), \\
	\tilde W(R,k) &= \exp \left(- \frac{(kR)^2}{2}\right),
\end{align}

\noindent \emph{k-space top-hat window function:}
\begin{align}
	W(R,x) &= \frac{1}{2\pi^2 R^3} \left( \frac{\sin(xR^{-1}) - xR^{-1} \cos(xR^{-1})}{(xR^{-1})^3} \right), \\[.5em]
	\tilde W(R,k) &= \Theta(R^{-1} -k ),
\end{align}
where $\Theta(x)$ is the Heaviside step function.
Note that, in the case of the real-space top-hat window and the Gaussian window, the normalizations of $W(R,x)$ are determined to satisfy $\int \dd^3 x W(R,x) = 1$.
On the other hand, in the case of the k-space top-hat window, since $\int \dd^3 x W(R,x)$ does not converge, the normalization of $W(R,x)$ is determined by the normalization of $\tilde W(R,k) (=\Theta(R^{-1}-k))$.
Then,  all the three window functions given here satisfy $\tilde W(R,k=0)=1$.

\begin{figure}
	\centering
	\includegraphics[width=.45\textwidth]{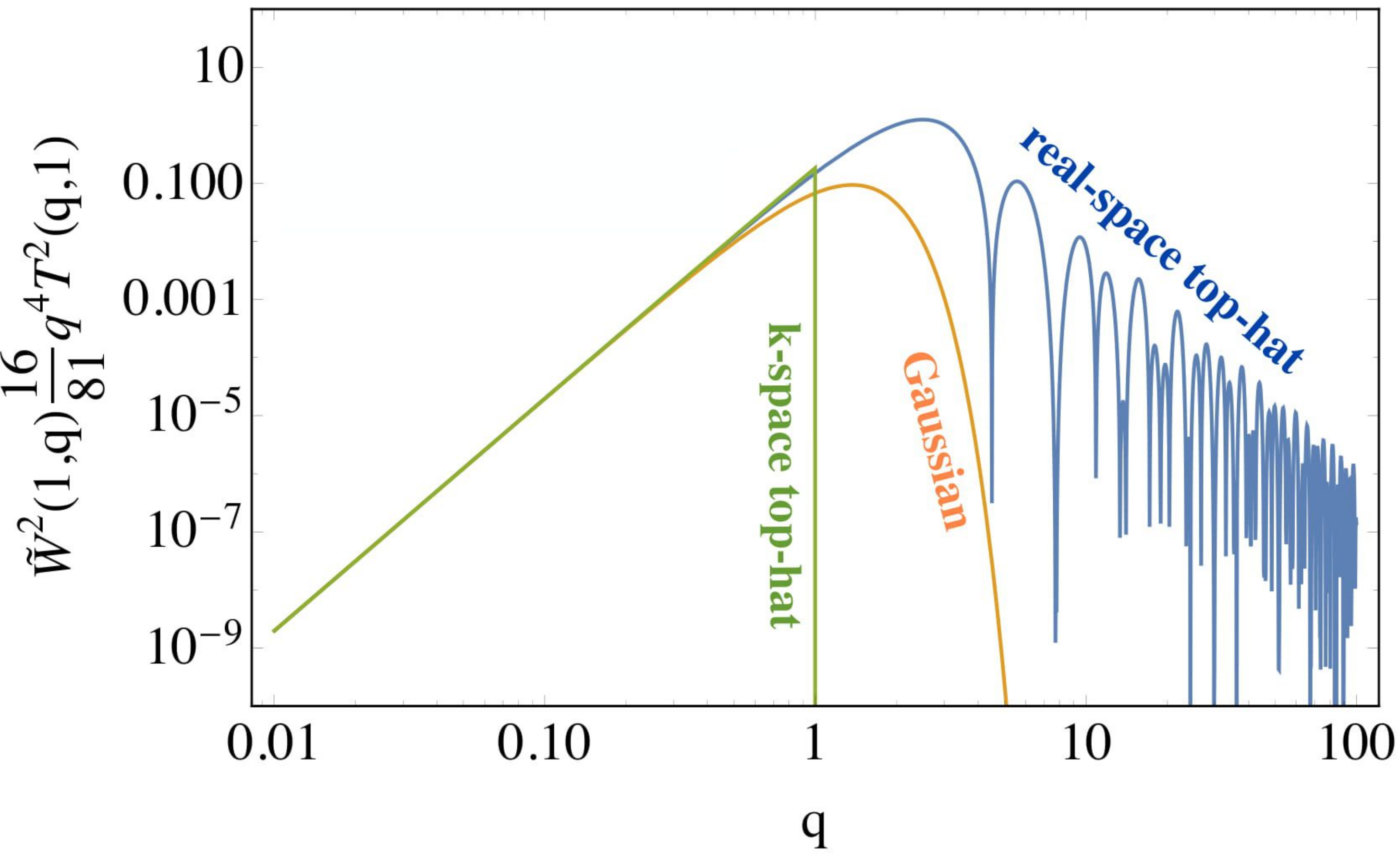}
	\caption{\small
	The integrands of Eq.~(\ref{eq:sigma}) with $k=1$ and a scale invariant power spectrum ($A_s = 1$).
	A blue line shows the case with the real-space top-hat window function.
	An orange line shows the case with the Gaussian window function.
	A green line shows the case with the k-space top-hat window function.
	}
	\label{fig:window_summary}
\end{figure}

To see the window function dependence, we consider a scale-invariant power spectrum of the curvature perturbations, $\mathcal P_\mathcal R(k) = A_s$, as a simple toy example.
Figure~\ref{fig:window_summary} shows the integrand of Eq.~(\ref{eq:sigma}) in this toy example with each window function, where $A_s$ is normalized as $A_s = 1$ in this figure.
The integrand with the real-space top-hat window function extends to the small scale ($q>1$).
On the other hand, the integrand with the k-space top-hat window function has a cutoff at $q=1$.
The case with the Gaussian window function corresponds to the intermediate case.
Doing the integral in Eq.~(\ref{eq:sigma}), we can derive the relation between $\sigma^2(M(k))$ and $A_s$ with each window function as
\begin{align}
	  \sigma^2(M(k)) = \begin{cases}
    1.06 \,A_s & (\text{real-space top-hat}) \\
    0.0867 \,A_s & (\text{Gaussian}) \\
    0.0472 \,A_s & (\text{k-space top-hat}).
  \end{cases}
\end{align}
The difference of the coefficients in front of $A_s$ is due to the difference of the integrands on the small scale ($q>1$).

%%%%%%%%%FIGURE%%%%%%%%%
%%%%%%%%%FIGURE%%%%%%%%%

Here, let us mention the uncertainties in the relation between the smoothing scale and PBH mass.
The volumes of the real-space top-hat and the Gaussian window functions are related to the normalized window functions as~\cite{Mo:2049511,Liddle:452061}
\begin{align}
	  W(R,x) = \begin{cases}
    \frac{1}{V(R)}\, \Theta( R-x) & (\text{real-space top-hat}) \\
    \frac{1}{V(R)}\, \exp\left(-\frac{x^2}{2R^2} \right) & (\text{Gaussian}).
  \end{cases}
  \label{eq:volume}
\end{align}
Then we get the volumes as
\begin{align}
	  V(R) = \begin{cases}
    4\pi R^3/3 & (\text{real-space top-hat}) \\
    (2\pi)^{3/2} R^3 & (\text{Gaussian}).
  \end{cases}
  \label{eq:volume_2}
\end{align}
On the other hand, it is not straightforward to define the volume of the k-space top-hat window function because $\int \dd^3 x W(R,x)$ diverges.
One way to define the volume is to use the relation $W(R,0) V(R) = 1$, which is satisfied in the case of the real-space top-hat and Gaussian window functions.
Following this prescription, we can define the volume as~\cite{Mo:2049511}
\begin{align}
	  V(R) =
    6 \pi^2 R^3 \quad (\text{k-space top-hat}).
  \label{eq:volume_3}
\end{align}
In the context of the halo formation, it is conventional to relate a mass included in an overdense region to a smoothing scale as $M(R) = V(R) \bar \rho$,
where $\bar \rho$ is the mean mass density at some given time~\cite{Mo:2049511,Liddle:452061}.
On the other hand, in the context of the PBH formation, it is conventional to identify the horizon scale with the smoothing scale and define the mass of the PBH as Eq.~(\ref{eq:pbhmass in k}) regardless of the window functions~\cite{Young:2014ana,Green:2004wb,Josan:2009qn,Sasaki:2018dmp}.
In this paper, we follow the convention of the PBH formation and assume that the value of $\gamma$ does not depend on the choice of window function for simplicity 
because there is no study discussing how much the relation between PBH mass and corresponding scale depends on the choice of window function so far.

Finally, let us summarize the features of each window function.
The real-space top-hat window function is often used because the relation between the mass of objects and the smoothing scale is unambiguously determined.
However, as we can see in Fig.~\ref{fig:window_summary}, the coarse-grained density perturbations are sensitive to the modes well inside the horizon
 and the careful treatments about the subhorizon evolution of the perturbations are needed, such as multiplying the transfer function given by Eq.~(\ref{eq:transfer_formula}).
The Gaussian window function is also often used because it is easy to handle analytically in both the real space and the Fourier space.
The k-space top-hat window function is used in the re-derivation of the Press-Schechter mass function~\cite{1990MNRAS.243..133P,1991ApJ...379..440B}.
This is because if we use the k-space top-hat window function, the trajectories of the density perturbations versus the smoothing scales are true Brownian random walk and become easy to treat.
However, in the case of this window function, there is an ambiguity on how to define the mass of the object with a given smoothing scale.

%%%%%%%%%%%%%%%%%%%%%%%%%%%%%%%%%%%%
\section{Constraints on PBH abundance}
\label{sec:constraints}
%%%%%%%%%%%%%%%%%%%%%%%%%%%%%%%%%%%%

In this section, we describe the constraints on the power spectra of the perturbations producing the $30M_\odot$ PBHs.
According to Ref.~\cite{Sasaki:2016jop}, if $f(30M_\odot) \sim \mathcal O(10^{-3})$,
PBHs can explain the merger rate expected by LIGO-Virgo Collaboration ($12$-$213$\,Gpc$^{-3}$yr$^{-1}$~\cite{Abbott:2017vtc}).
Substituting $f(30M_\odot) =10^{-3}$ into Eq.~(\ref{eq:frac1}), we can estimate $\beta(30M_\odot) \simeq \mathcal O(10^{-11})$ for LIGO PBHs.
From Eq.~(\ref{eq:sigma}), we can also estimate the power spectrum required to produce LIGO PBHs as $\mathcal P_\mathcal R(10^{6}\text{Mpc}^{-1}) \sim \mathcal O(0.01)$.
Since the perturbations for LIGO PBHs are large, we cannot neglect the second order effect of the perturbations,
 which produces $\mu$-distortion~\cite{Fixsen:1996nj,Chluba:2012we,Kohri:2014lza} and the stochastic GWs constrained by PTA experiments~\cite{Saito:2008jc,Saito:2009jt,Bugaev:2009zh,Bugaev:2010bb}.\footnote{
In addition to the $\mu$-distortion and PTA constraints, there are constraints from the current abundance of light elements as
$\mathcal P_\mathcal R < \mathcal O (0.01)$ on $10^4$Mpc$^{-1} \lesssim k \lesssim 10^5$Mpc$^{-1}$ ~\cite{Jeong:2014gna,Nakama:2014vla,Inomata:2016uip}.
Although the constrained scale is a little smaller than the scale constrained by the $\mu$-distortion,
the constraint is weak and has some uncertainties compared to the $\mu$-distortion constraints.
Hence, we neglect the constraints from the current abundance of light elements in this paper.
}

%%%%%%%%%%%%%%%%%%%%%%%%%%
\paragraph*{\bf $\mu$-distortion.}
%%%%%%%%%%%%%%%%%%%%%%%%%%

Small scale perturbations dissipate through the photon diffusion, which distorts the CMB spectrum and deviates it from the Planck distribution.
The relation between the perturbation scale and the redshift when the diffusion of the perturbations occur most efficiently is numerically calculated as~\cite{Chluba:2012we}
\begin{align}
	z_\text{peak} \simeq 4.5 \times 10^5 \left( \frac{k}{10^3 \text{Mpc}^{-1}} \right)^{2/3}.
	\label{eq:z_peak}
\end{align}
From Eq.~(\ref{eq:z_peak}), we can see that the perturbations on $50$\,Mpc$^{-1}$ < k < $10^{4}$\,Mpc$^{-1}$ are diffused after the turn-off of the double Compton scattering interaction ($z \sim 2\times 10^{6}$~\cite{Hu:1992dc}), but before the turn-off of the Compton scattering interaction ($z \sim 5 \times 10^{4}$~\cite{Hu:1992dc}).
During this phase, while the diffusion of perturbations injects energy from perturbations to the background and the photon distribution goes to kinetic equilibrium due to the Compton scattering, the number of photons remains constant because there is no number changing interaction such as the double Compton scattering.
This is the reason why the perturbations on $50$\,Mpc$^{-1}$ < k < $10^{4}$\,Mpc$^{-1}$ make the photon distribution follow the Bose distribution with a finite chemical potential.
The parameter of $\mu$-distortion, $\mu$, is defined as
\begin{align}
	f = \frac{1}{\ee^{\frac{p}{T} - \mu} -1},
\end{align}
where $f$ is the CMB photon distribution and $p$ and $T$ are the photon momentum and temperature.
This $\mu$ parameter is constrained by COBE/FIRAS as~\cite{Fixsen:1996nj}
\begin{align}
	|\mu| < 9 \times 10^{-5}.
\end{align}
According to Ref.~\cite{Chluba:2012we}, as for the monochromatic power spectra defined as $\mathcal P_\mathcal R(k) = A \delta( \text{log} k - \text{log} k_* )$,
 the relation between the $\mu$ parameter and the amplitude $A$ is given by
\begin{align}
	\mu \simeq 2.2 A \left[ \exp\left( - \frac{\hat k_*}{5400}\right) - \exp \left( - \left[ \frac{\hat k_*}{31.6} \right]^2 \right) \right],
	\label{eq:mu_power_mono}
\end{align}
where $\hat k = k$\,Mpc.
As for general curvature power spectra, the $\mu$ parameter is given by
\begin{align}
	\mu \simeq 2.2 \int^{\infty}_{k_\text{min}} \mathcal P_\mathcal R (k) \left[ \exp\left( - \frac{\hat k}{5400}\right) - \exp \left( - \left[ \frac{\hat k}{31.6} \right]^2 \right) \right] \dd\, \text{ln} k,
	\label{eq:mu_power}
\end{align}
where $k_\text{min}\simeq 1$\,Mpc$^{-1}$.

%%%%%%%%%%%%%%%%%%%%%%%%%%
\vspace{.5em}
\paragraph*{\bf Stochastic GWs from second order.}
%%%%%%%%%%%%%%%%%%%%%%%%%%

Although GWs (tensor perturbations) are not produced by linear curvature (scalar) perturbations,
GWs are produced by the second-order curvature perturbations.
In this subsection, we briefly review the formula for the stochastic GWs induced by the second order curvature perturbations during the radiation-dominated era (see e.g. Refs.~\cite{Saito:2008jc,Saito:2009jt,Bugaev:2009zh,Bugaev:2010bb,Inomata:2016rbd} for details).

When the perturbations cross the sound horizon, the stochastic GWs are efficiently induced by the source terms from the second order perturbations.
After the perturbations renter the horizon, the source terms are irrelevant to the GWs and the induced GWs behave as radiation without any sources; that is, the energy density of the induced GWs is proportional to $a^{-4}$.
Therefore, the density parameter of the induced GWs is given by
\begin{align}
	\Omega_\text{GW}(\eta_0, k) h^2 &= \left(\frac{a_c^2 H_c}{a_0^2 H_0} \right)^2 \Omega_\text{GW}(\eta_c,k) h^2 \nonumber \\
	&= \left(\frac{g_{*s,0}}{g_{*s,c}} \right)^{4/3} \frac{g_{*,c}}{g_{*,0}} \,\Omega_{r,0} h^2 \Omega_\text{GW}(\eta_c,k) \nonumber \\
	&= 0.83 \left(\frac{g_c}{10.75}\right)^{-1/3} \Omega_{r,0} h^2 \Omega_\text{GW}(\eta_c,k),
	\label{eq:omega_gw_form}
\end{align}
where $\Omega_\text{GW}(\eta_0,k)$ is the differential density parameter related to the total density parameter as
\begin{align}
	\Omega_\text{GW}(\eta_0) = \int \dd \, \text{ln} k\, \Omega_\text{GW} (\eta_0,k).
\end{align}
$\eta_c$ (before the matter-radiation equality time $\eta_\text{eq}$) represents the conformal time when the induced GWs start to behave as radiation without any source.
$a$, $H$ and $g_{*s}$ are the scale factor, the Hubble parameter and the effective number of relativistic degrees of freedom contributing to the entropy density.
The subscripts "$0$" and "$c$" mean the values at the present and $\eta_c$ respectively.
We assume that the PBH is produced before the electron annihilation ($T>0.1$MeV) and $g_{*,c}=g_{*s,c}$ is satisfied.
To derive the third line of Eq.~(\ref{eq:omega_gw_form}), we have substituted $g_{*,0} = 3.36$ and $g_{*s,0}=3.91$~\cite{Kolb:206230}.
We take $g_{*,c} = 10.75$ for LIGO PBHs.
$\Omega_{r,0} h^2 (\simeq 4.2\times 10^{-5})$ is the current density parameter of radiation.\footnote{
In this paper, we assume neutrinos are massless.
If we assume the neutrinos behave as non-relativistic matter at the present,
Eq.~(\ref{eq:omega_gw_form}) should be modified as
\begin{align}
	\Omega_\text{GW}(\eta_0, k) h^2 &= 1.4 \left(\frac{g_c}{10.75}\right)^{-1/3} \Omega_{r,0} h^2 \Omega_\text{GW}(\eta_c,k), \nonumber
\end{align}
where $\Omega_{r,0} h^2 = 2.5 \times 10^{-5}$.
We can see that the factor in front of $\Omega_\text{GW}(\eta_c,k)$ is the same, even if we assume neutrinos are non-relativistic at the present.
}
The density parameter at $\eta_c$ is given by~\cite{Inomata:2016rbd}
\begin{align}
	\Omega_\text{GW}(\eta_c,k) = \frac{8}{243} \int^{\infty}_0 \dd v &\int^{1+v}_{|1-v|} \dd u
	\left[ \frac{4v^2 - (1-u^2 + v^2 )^2 }{4vu} \right]^2  \nonumber \\
	& \times \mathcal P_\mathcal R(kv) \mathcal P_\mathcal R(ku) \overline{I^2 (v,u,k \eta_c)},
	\label{eq:gw_power}
\end{align}
where the overline represents the time average over the oscillations.
$I(v,u,x)$ is defined as
\begin{align}
	I(v,u,x) \equiv \int^x_0 \dd &\bar x \bar x \sin(x- \bar x) \left[ 3 \Psi(v \bar x) \Psi(u \bar x) \right. \nonumber \\
	&+ \bar x \left\{ \Psi(v\bar x) u \Psi'(u \bar x) + v \Psi'(v \bar x) \Psi(u \bar x) \right\} \nonumber \\
	& \qquad \qquad\qquad \left. + \bar x^2 u v \Psi'(u \bar x) \Psi'(v \bar x) \right],
\end{align}
where $\Psi(x)$ is given by
\begin{align}
	\Psi(x) = \frac{9}{x^2} \left[ \frac{\sin(x/\sqrt{3})}{x/\sqrt{3}} - \cos( x/\sqrt{3}) \right].
\end{align}
The frequency of the induced GWs corresponds to the scale of the perturbations producing PBHs.
In the case of $30 M_\odot$ PBHs, the frequency of the induced GWs is $f \sim$nHz (see Eq.~(\ref{eq:pbhmass_freq})),
which is close to the detectable frequency of the PTA experiments ($f \gtrsim$ nHz )~\cite{Lentati:2015qwp,Shannon:2015ect,Arzoumanian:2015liz}.

%%%%%%%%%%%%%%%%%%%%%%%%%%%%%%%%%%%%
\section{Sharpness of power spectrum}
\label{sec:sharpness}
%%%%%%%%%%%%%%%%%%%%%%%%%%%%%%%%%%%%

In the previous section, we have showed that the perturbations producing $30M_\odot$ PBHs ($k\sim 10^6$Mpc$^{-1}$) are constrained by
$\mu$-distortion observations from the larger scale ($k<10^4$Mpc$^{-1}$)
 and PTA observations from the smaller scale ($k>10^6$Mpc$^{-1}$ or $f>$nHz ).
Hence, the power spectrum must have a peak at $k\sim 10^6$Mpc$^{-1}$.
We show that the required sharpness of the peak depends on the choice of a window function.

In this section, we investigate how sharp the power spectrum should be in each window function.
In order to discuss this issue quantitatively, we parametrize the shape of the power spectrum around the peak scale as
\begin{align}
	  \mathcal P_{\mathcal R, \text{peak}} (k) = \begin{cases}
    A_* \left( \frac{k}{k_*} \right)^x & (k<k_*) \\[.5em]
    A_* \left( \frac{k}{k_*} \right)^{-y} & (k>k_*) ,
  \end{cases}
  \label{eq:def_power}
\end{align}
where $k_*$ and $A_*$ are the pivot scale and the amplitude of the power spectrum at the pivot scale.
The parameters $x$ and $y$ indicate the tilts of the spectrum.
We take $0<x<8$ and $0<y<8$ as concrete values.
Since the power spectrum on the large scale ($k<1$Mpc$^{-1}$) is determined by the CMB and the large scale structure (LSS) observations as $\mathcal P_\mathcal R \simeq 2\times 10^{-9}$~\cite{nicholson2009reconstruction,nicholson2010reconstruction, bird2011minimally},
we assume this parametrization is valid only on the small scale ($k>1$Mpc$^{-1}$).
We take the shape of the power spectrum on all scales as
\begin{align}
	\mathcal P_\mathcal R(k) = \begin{cases}
	2\times 10^{-9} & (k<k_c) \\[.5em]
	\mathcal P_{\mathcal R, \text{peak}} (k) & (k>k_c),
  \end{cases}
\end{align}
where
$k_c$ is defined as the scale on which $\mathcal P_{\mathcal R, \text{peak}} (k_c) = 2 \times 10^{-9}$ and
the tilt of the power spectrum on large scales is neglected for simplicity.
To be consistent with the observations on large scales,
the parametrized power spectrum must be
\begin{align}
	&k_c > 1\text{Mpc}^{-1} \nonumber \\
	\Rightarrow \quad &
	A_*\left( \frac{1 \text{Mpc}^{-1}}{k_*} \right)^x < 2\times 10^{-9}.
	\label{eq:cmb_constraint}
\end{align}

Here, we explain the procedure.
First, we input the values of $x$ and $y$ and search for $k_*$ and $A_*$ in which $f(M)$ has its maximum at $30M_\odot$\footnote{
Although LIGO-Virgo has detected several light BHs with $\sim10M_\odot$ so far~\cite{Abbott:2016nmj},
we focus on only the $30M_\odot$ BHs in this paper.
} and the maximum height is $f(30M_\odot) = 10^{-3}$~\cite{Sasaki:2016jop}.
Next, using Eq.~(\ref{eq:def_power}) and the derived parameter sets, $(x, y, k_*, A_*)$, we calculate the $\mu$ parameter with Eq.~(\ref{eq:mu_power}) and the stochastic GWs with Eq.~(\ref{eq:gw_power}).
Comparing the resultant values of $\mu$ and $\Omega_\text{GW}h^2$ with the observational constraints,
we check whether or not the input $x$ and $y$ are consistent with the observations.
In addition, we check whether the derived parameter sets, $(x, y, k_*, A_*)$, are consistent with Eq.~(\ref{eq:cmb_constraint}).
Then, we change the input $x$ and $y$ and repeat the above steps.
Finally, we derive the allowed parameter region of $x$ and $y$ in which $30M_\odot$ PBHs for LIGO events can exist without contradicting the observations.

To take into account the other uncertainties of the PBH formation than those originating from the choice of a window function, 
we take the values of $\gamma$ as
\vspace{-.3em}
\begin{center}
case (i): $\gamma = 0.2$,

\vspace{.3em}
case (ii): $\gamma = 1$.

\end{center}

%%%%%%%%%FIGURE%%%%%%%%%
\begin{figure}
	\centering
	\includegraphics[width=.45\textwidth]{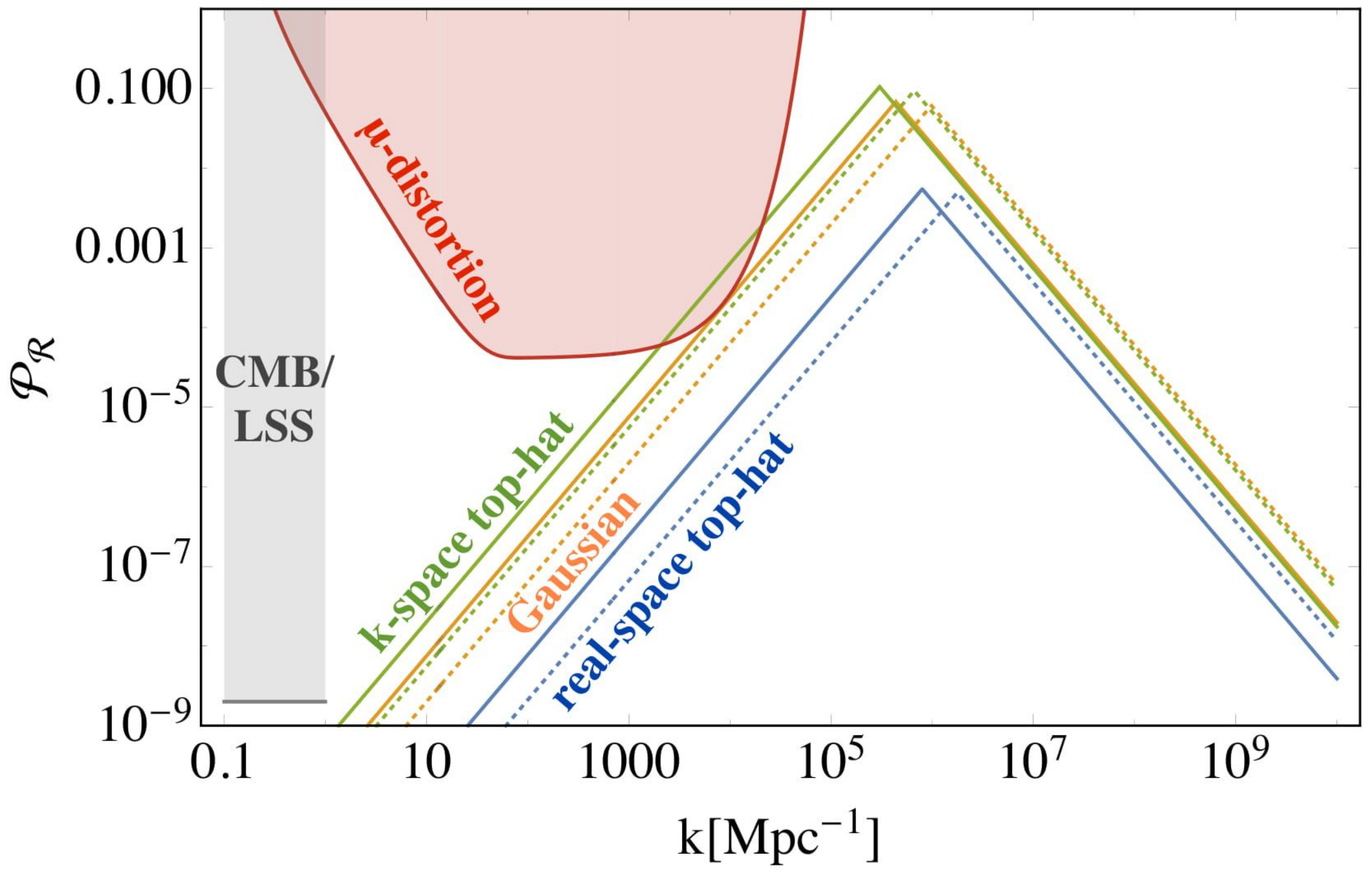}
	\caption{\small
	Power spectra of the curvature perturbations required for LIGO PBHs ($f(30M_\odot) \simeq 10^{-3}$) with each window function.
	In this figure, we take $x=1.5$ and $y=1.5$.
	The blue solid (dotted) line shows the power spectra with the real-space top-hat window function with $\gamma=0.2$ ($\gamma=1$).
	The orange solid (dotted) line shows the power spectra with the Gaussian window function with $\gamma=0.2$ ($\gamma=1$).
	The green solid (dotted) line shows the power spectra with the k-space top-hat window function with $\gamma=0.2$ ($\gamma=1$).
	The gray shaded region is excluded by the CMB and LSS observations, which are given by Eq.~(\ref{eq:cmb_constraint}). 
	For comparison, we show the constraints from $\mu$-distortion observations by COBE/FIRAS ($|\mu|<9\times10^{-5}$~\cite{Fixsen:1996nj})
	 in the case of the monochromatic power spectrum with the red shaded region, which is given by Eq.~(\ref{eq:mu_power_mono}).
	 Note that since the power spectra we consider are not monochromatic functions,
	 the intersection between the lines and the red shaded region does not necessarily mean an inconsistency with the observations and vice versa.	
	}
	\label{fig:curvature_revised}
\end{figure}
%%%%%%%%%FIGURE%%%%%%%%%

%%%%%%%%%FIGURE%%%%%%%%%
\begin{figure}
	\centering
	\includegraphics[width=.45\textwidth]{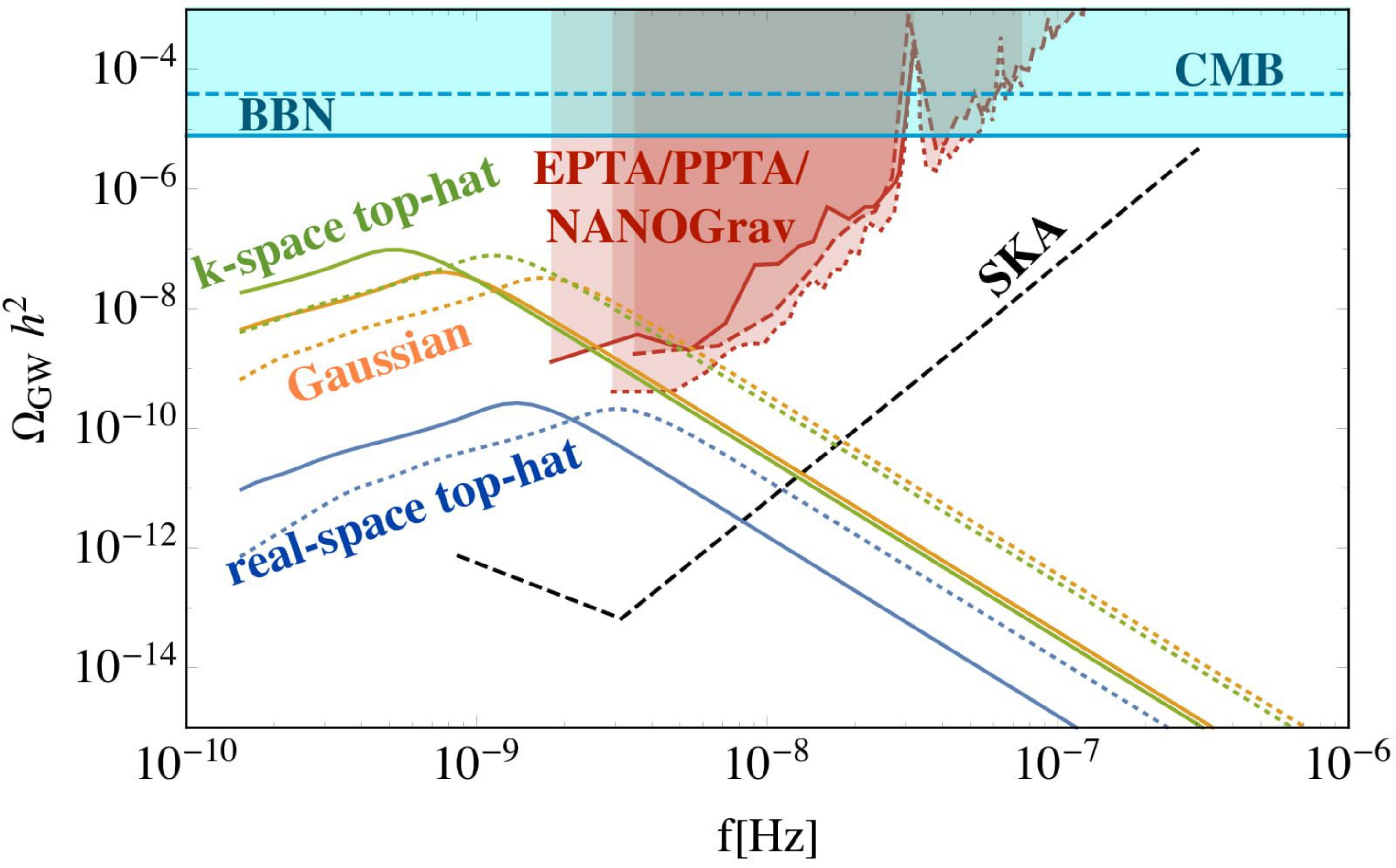}
	\caption{\small
	Energy density parameters of the GWs induced by the curvature perturbations shown in Fig.~\ref{fig:curvature_revised}.
	The colors of lines in this figure correspond to the colors of lines in Fig.~\ref{fig:curvature_revised}.
	For example, the blue solid line shows the GWs induced by the curvature perturbations plotted in Fig.~\ref{fig:curvature_revised} with the blue solid line.
	The red shaded region is excluded by the current PTA observations~\cite{Lentati:2015qwp,Shannon:2015ect,Arzoumanian:2015liz}.
	The cyan shaded region is excluded by the current big bang nucleosynthesis observations~\cite{Pierpaoli:2003kw,Smith:2006nka} and the cyan dashed line shows the upper bound on the GW density parameter from the CMB observations~\cite{Smith:2006nka}.
	A black dashed line shows the future prospects of SKA~\cite{Moore:2014lga,Janssen:2014dka}.
	}
	\label{fig:gw_revised}
\end{figure}
%%%%%%%%%FIGURE%%%%%%%%%

%%%%%%%%%%%%%%%%%%%%%%%%%%
\paragraph*{\bf Concrete examples.}
%%%%%%%%%%%%%%%%%%%%%%%%%%

Before we discuss the main results of the allowed parameter region of $x$ and $y$, we consider an example with concrete values, $x=1.5$ and $y=1.5$, in order to show the part of the procedure.
 Figures~\ref{fig:curvature_revised} and \ref{fig:gw_revised} show the power spectra of the curvature perturbations with $x=1.5$ and $y=1.5$ for LIGO PBHs and the induced stochastic GWs. 

In Fig.~\ref{fig:curvature_revised}, we can see that the values of $k_*$ and $A_*$ depend on the choice of a window function.
This is because the contribution from the subhorizon modes depends on the window function.
For example, in the case of the real-space top-hat window function, the subhorizon modes have a large contribution to the coarse-grained density perturbations compared to the case with the other window functions (see Fig.~\ref{fig:window_summary}) and therefore $k_*$ is large and $A_*$ is small.
In Fig.~\ref{fig:curvature_revised}, for comparison, we also show the $\mu$-distortion constraints on the monochromatic power spectra with a red shaded region using Eq.~(\ref{eq:mu_power_mono}).
Although, in the case of non-monochromatic power spectra, the intersection between the lines and the red shaded region does not necessarily mean an inconsistency with the observations and vice versa,
the intersection can be used for a rough estimate of the consistency with $\mu$-distortion observation even in the case.
For example, as for the solid lines in Fig.~\ref{fig:curvature_revised} (case (i)), the $\mu$ parameters are $\mu = 6.0 \times 10^{-6}$ for the real-space top-hat window function (blue solid),
$\mu = 1.8 \times 10^{-4}$ for the Gaussian window function (orange solid),
and $\mu = 4.9 \times 10^{-4}$ for the k-space top-hat window function (green solid).
In this case, only the result with the real-space top-hat window function is consistent with the $\mu$-distortion observation by COBE/FIRAS ($|\mu|< 9\times 10^{-5}$)~\cite{Fixsen:1996nj}.
In Fig.~\ref{fig:curvature_revised}, we can also find that the peak scales of the curvature perturbations in case (ii) are smaller than those in case (i) 
because the relation between the PBH mass and the perturbation scale depends on the value of $\gamma$ as Eq.~(\ref{eq:pbhmass in k}).

In Fig.~\ref{fig:gw_revised}, we plot the PTA constraints with a red shaded region.
Unlike in the case of the $\mu$-distortion constraints, the intersection between the lines of the induced stochastic GWs and the red shaded region means that the prediction of induced GWs is inconsistent with the observations.\footnote{
If the frequency dependence of the stochastic GWs is $\Omega_\text{GW} h^2 \propto f^{2/3}$, which is predicted from the supermassive BH binaries, the PTA constraints could possibly become more severe~\cite{Lentati:2015qwp,Shannon:2015ect,Arzoumanian:2015liz}.
In this paper, since we consider the peak-like GW spectra, we take the results of the general spectra.
}
In particular, in case (ii), the peak scale is almost in the observable range of the PTA and therefore the PTA constraints become more severe than those in case (i) (see the results in next subsection).

From Figs.~\ref{fig:curvature_revised} and~\ref{fig:gw_revised}, we can see that the required amplitude of the power spectrum is smallest in the case of the real-space top-hat window function 
and hence the induced GW is smallest too.
This means that the perturbations with the real-space top-hat window function can avoid the constraints more easily than those with the other window functions.

%%%%%%%%%%%%%%%%%%%%%%%%%%
\paragraph*{\bf Results.}
%%%%%%%%%%%%%%%%%%%%%%%%%%

\begin{figure}
	\centering
	\includegraphics[width=.45\textwidth]{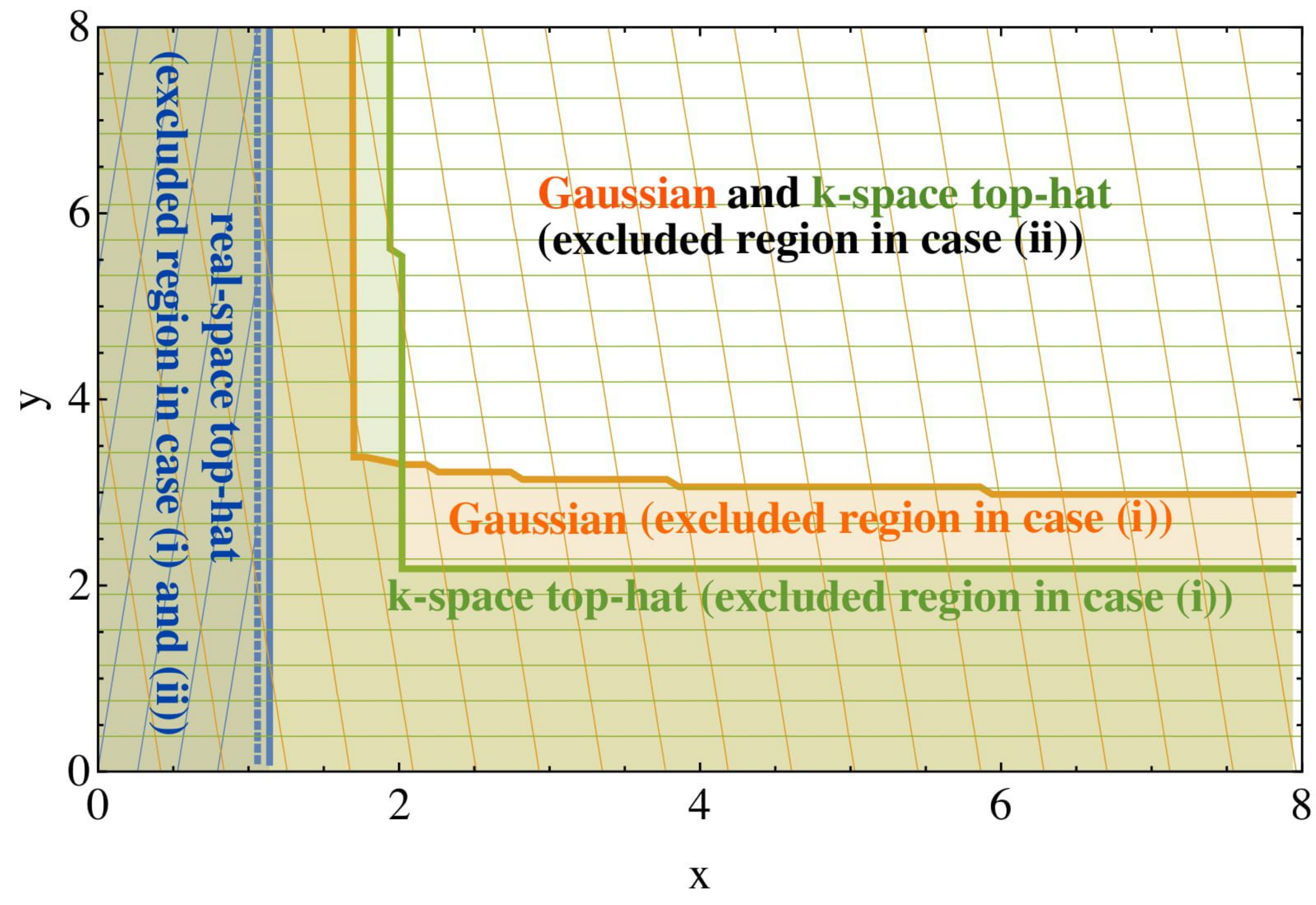}
	\caption{\small
	Constraints on the sharpness of the power spectra of the curvature perturbations (see Eq.~(\ref{eq:def_power}) for the definition of $x$ and $y$).
	Each color corresponds to each window function, blue: real-space top-hat window, orange: Gaussian window, and green: k-space top-hat window.	
	Case (i): the shaded regions show the regions excluded by the observations.
	The solid lines show the boundaries of the excluded regions in case (i).
	Case (ii): the hatched regions show the regions excluded by the observations.
	The (blue) dotted line shows the boundary of the excluded region in case (ii).
	Note that, in the case of the Gaussian and k-space top-hat window function, all of the region is excluded by the observations.
	(Hence, orange or green dotted lines are not plotted in this figure.)
	}
	\label{fig:xy_const_both}
\end{figure}

Now, let us discuss the main results of the allowed region of $x$ and $y$.
Figure~\ref{fig:xy_const_both} shows the summary of the constraints on the sharpness of the power spectra of the curvature perturbations.
The lower bounds on $x$ and $y$ come from the $\mu$-distortion/CMB anisotropy observations and the PTA observations respectively.
Note that, in Fig.~\ref{fig:xy_const_both}, there is no lower bound on $y$ for the real-space top-hat window function.
This is because the induced GWs at the peak frequency in both case (i) and (ii) are smaller than the upper limits from the PTA observations.
Therefore, if the real-space top-hat window function is taken, the power spectrum needs no suppression on the smaller scale side of the peak scale and therefore the PBH mass spectrum could extend from $30M_\odot$ to the lighter mass range.
On the other hand, in the case of the Gaussian and the k-space top-hat window functions, the value of $y$ is constrained by the PTA observations.
In particular, in case (ii), the PTA observations severely constrain the value of $y$ and exclude all the parameter region ($0<x<8$ and $0<y<8$).

Finally, let us summarize the constraints on $x$ and $y$,

\noindent Case (i):
\begin{align}
    x \gtrsim 1.1  & \qquad \qquad \quad \ \  (\text{real-space top-hat window}), \nonumber \\[.5em]
    x \gtrsim 1.7&, y \gtrsim 3.1  \qquad (\text{Gaussian window}), \nonumber \\[.5em]
    x \gtrsim 2.0&, y \gtrsim 2.2  \qquad (\text{k-space top-hat window}), \nonumber
\end{align}

\noindent Case (ii):
\begin{align}
    &x \gtrsim 1.0   \qquad \qquad \quad \ \  (\text{real-space top-hat window}), \nonumber \\[.5em]
    &\text{no allowed region} \qquad (\text{Gaussian window}), \nonumber \\[.5em]
    &\text{no allowed region}  \qquad (\text{k-space top-hat window}).\nonumber
\end{align}
From these results, we can see that $x\gtrsim 1$ is needed for any window functions.
This means that a single field slow-roll inflation is not likely to be appropriate for the LIGO PBHs 
because, during the usual slow-roll period, the tilt of curvature perturbations $n_s$ is described with the slow-roll parameters ($\epsilon, \eta\, (\ll 1)$) as $n_s = 1-6\epsilon + 2\eta$.
The $n_s$ corresponds to the $x$ and $y$ as $n_s -1 = x \text{ or } y$ and it is difficult to achieve $x>1$ during the usual slow-roll period.
Instead, inflation models with multiple fields~\cite{GarciaBellido:1996qt,Inomata:2016rbd, Ando:2017veq} and inflation models that violate the usual slow-roll conditions~\cite{Garcia-Bellido:2017mdw, Ezquiaga:2017fvi, Germani:2017bcs, Kannike:2017bxn, Ballesteros:2017fsr, Hertzberg:2017dkh,Cicoli:2018asa} are favored.

%%%%%%%%%FIGURE%%%%%%%%%
%%%%%%%%%FIGURE%%%%%%%%%

%%%%%%%%%%%%%%%%%%%%%%%%%%%%%%%%%%%%
\section{Conclusions}
\label{sec:conclusions}
%%%%%%%%%%%%%%%%%%%%%%%%%%%%%%%%%%%%

LIGO-Virgo Collaboration has detected $30M_\odot$ BHs and PBHs are good candidates for such BHs.
PBHs can be produced by the large curvature perturbations that exit the horizon during the inflation era and reenter the horizon during the radiation-dominated era.
The power spectrum of the curvature perturbations for LIGO PBH formation has a peak at $k\sim 10^{6}$Mpc$^{-1}$ and its amplitude is $\mathcal P_\mathcal R \sim \mathcal O(0.01)$.
Such power spectra are severely constrained by the $\mu$-distortion and PTA observations.
Therefore, when we discuss the PBHs for LIGO events, it is important to investigate the relation between the PBH abundance and the power spectrum.

While we can observe the mass spectra of BHs through GW detections,
inflation models predict the power spectrum of the curvature perturbations.
There are some uncertainties in the formulae that relate the power spectrum to the mass spectrum of PBHs.
In particular, since the PBH abundance depends on the PDF of the density perturbations in real space,
we must calculate the coarse-grained density perturbations with a window function.
However, it is non-trivial what window function should be used.
Although the PBH abundance should have a one-to-one correspondence with the power spectra in realistic situations,
there are uncertainties on the relation between the PBH abundance and the power spectra due to the uncertainties on the choice of a window function.
In this paper, we have investigated how much the uncertainties on the choice of a window function affect the power spectra required for LIGO PBHs and the induced observable quantities such as the $\mu$-distortion and stochastic GWs.

As a result, we have found that the uncertainties on the choice of a window function lead to large uncertainties on the power spectrum required for LIGO PBHs and the induced observable quantities.
In particular, if we take the real-space top-hat window function, 
there are no constraints from the PTA observations on the power spectra.
This means that there is a possibility of the PBH mass spectra extending from $30M_\odot$ to the lighter mass range.
In other words, when we discuss the possibility of the GW detection by the PTA in the context of LIGO PBHs, we should take care of the uncertainties on the choice of the window functions.

Finally, let us mention some uncertainties that we do not take into account in this paper.
First, as we mentioned in Sec.~\ref{sec:window}, there are uncertainties on the relation between the PBH mass and the smoothing scale.
If we take the convention used in the study of the halo formation unlike the main body of this paper, the PTA constraints become relatively severe in the case of the Gaussian and the k-space top-hat window function.
This is because the PBH with a given mass should be produced by the smaller scale perturbations.
Second, throughout this paper, we have assumed that if $f(30M_\odot) \sim \mathcal O(10^{-3})$ is satisfied, PBHs can explain the LIGO events~\cite{Sasaki:2016jop,Ali-Haimoud:2017rtz}.
However, the results in Refs.~\cite{Sasaki:2016jop,Ali-Haimoud:2017rtz} are based on the assumption of the monochromatic mass function of PBHs.
In the case of the broad mass spectrum, which corresponds to the case with small $x$ and $y$ in Eq.~(\ref{eq:def_power}), the relation between the PBH abundance and the expected merger rate could possibly be modified~\cite{Ali-Haimoud:2017rtz}.\footnote{
Although the result strongly depends on the properties of the PBH clustering, which have some uncertainties, the merger rate in the case of the broad mass spectrum is calculated in Refs.~\cite{Raidal:2017mfl,Clesse:2017bsw}.
}
Third, although we have assumed the Gaussian PDF in this paper, the non-Gaussianity could change the results.
If we consider the non-Gaussianity, the relation between the PBH abundance and the required power spectrum changes~\cite{Byrnes:2012yx,Young:2013oia}.
Depending on the amount of the non-Gaussianity, the constraints from the PTA and the $\mu$-distortion could be significantly weakened~\cite{Nakama:2016kfq,Nakama:2016gzw,Nakama:2017xvq}.

%%%%%%%%%%%%%%%%%%%%%%%%%%%%%%%%%%
%%%%%%%%%%% Acknowledge %%%%%%%%%%%
%%%%%%%%%%%%%%%%%%%%%%%%%%%%%%%%%%
\section*{Acknowledgements}
{\small
\noindent
This work was supported by JSPS KAKENHI Grant Nos.~17H01131 (M.K.)
and 17K05434 (M.K.), MEXT KAKENHI Grant No.~15H05889 (M.K.),
World Premier International Research Center Initiative (WPI Initiative), MEXT, Japan
(K.A., K.I., M.K.),
and Advanced Leading Graduate Course for Photon Science (K.A., K.I.).
}

%%%%%%%%%%%%%%%%%%%%%%%%%%%%%%%%%
%%%%%%%%%%% References %%%%%%%%%%%
%%%%%%%%%%%%%%%%%%%%%%%%%%%%%%%%%
\small
%\bibliographystyle{apsrev4-1}
%\bibliography{pbh_and_window}

\end{document}